\documentclass[aps,twocolumn,groupedaddress,showpacs,floatfix]{revtex4}

\usepackage{amssymb}
\usepackage{graphicx}

\begin{document}

\title{RF spectroscopy in a resonant RF-dressed trap}
\author{R.~Kollengode Easwaran$^1$}
\author{L.~Longchambon$^1$}
\author{P.-E.~Pottie$^1$}
\author{V.~Lorent$^1$}
\author{H.~Perrin$^1$}\email{helene [dot] perrin [at] univ - paris13 [dot] fr}
\author{B.\,M.~Garraway$^2$}

\affiliation{$^1$Laboratoire de physique des lasers, Institut Galil\'ee, Universit\'e Paris 13 and CNRS,
Avenue J.-B. Cl\'ement, F-93430 Villetaneuse, France\\
$^2$Department of Physics and Astronomy, University of Sussex, Brighton BN1 9QH, United Kingdom}

\date{\today}

\begin{abstract}
We study the spectroscopy of atoms dressed by a resonant radiofrequency (RF) field inside an inhomogeneous magnetic field and confined in the resulting adiabatic potential. The spectroscopic probe is a second, weak, RF field. The observed line shape is related to the temperature of the trapped cloud. We demonstrate evaporative cooling of the RF-dressed atoms by sweeping the frequency of the second RF field around the Rabi frequency of the dressing field.
\end{abstract}


\maketitle

\section{Introduction}
Since the initial proposal of Zobay and Garraway~\cite{Zobay2001}, adiabatic potentials for ultracold atoms, resulting from the radiofrequency (RF) dressing of atoms in a inhomogeneous magnetic field, have been demonstrated~\cite{Colombe2004} and are used by many groups to propose and realize unconventional traps~\cite{Lesanovsky2006,Morizot2006,Courteille2006,Schumm2005,Extavour2006,Jo2007,Heathcote2009}, in particular in the framework of atom chips~\cite{Schumm2005,Extavour2006,Jo2007}. They offer the great advantage of a simple tailoring of the trap potential, allowing the realisation of a double well~\cite{Schumm2005} or a ring trap~\cite{Morizot2006,Heathcote2009} with dynamically adjustable parameters. The potential seen by the dressed atoms is due to the spatial variation of the RF coupling strength or the RF detuning, or both. These RF-dressed traps are usually loaded directly from a conventional magnetic trap, by ramping up the RF amplitude or sweeping the RF frequency.

Depending on the conditions, the adiabatic potential may result essentially from a light shift induced by a variation of the RF coupling, as was first demonstrated in a pioneering experiment at NIST with microwave radiation~\cite{Spreeuw1994}. In this case, denoted here as the `off-resonant configuration', the detuning of the RF field to the magnetic Zeeman splitting is negative everywhere and the atoms are expelled from the most resonant central region, leading for example to a double well structure well suited for matter wave interferometry\,\cite{Hofferberth2006}. On the other hand, the trapping potential may be due essentially to the variation of the detuning, the atoms being then attracted towards the regions of zero detuning~\cite{Zobay2001}. In this `resonant configuration', the atoms sit at points where the RF field is resonant with the magnetic level spacing, which makes them more sensitive to fluctuations of the RF frequency, and heating due to frequency fluctuations was indeed observed~\cite{Colombe2004,White2006,Morizot2008}. This heating may be avoided if a low noise RF source is used~\cite{Morizot2008}, leading to traps compatible with Bose-Einstein condensates. However, the transfer of the atoms from the initial magnetic trap to the RF-dressed trap is not usually adiabatic, which results in an initial heating in the loading phase and the destruction of the condensate~\cite{Colombe2004,White2006}.

This limitation motivated a theoretical study of evaporative cooling inside the dressed trap, by means of a second RF field~\cite{Garrido2006}. In the case of a non-resonant RF trap, this scheme was used to prepare independent condensates in a double well~\cite{Hofferberth2006}. In this paper, we present results of spectroscopy in the dressed trap, obtained by recording the atom losses induced by a very weak second RF field~\cite{Hofferberth2007,vanEs2008}. The shape of the resonance is directly linked to the energy distribution of the atoms in the dressed trap. We discuss the position of the expected lines and the strength of the second RF coupling. Finally, we discuss the choice of the second field frequency for evaporative cooling of the atoms dressed by the main RF field.

\section{Spectroscopy of the RF-dressed trap}
\label{sec:spectro}
The spectroscopy of an RF-dressed trap is performed in the following way: A weak probing field can induce transitions between dressed states, which results in a loss of atoms from the adiabatic potential. The losses are monitored as a function of the frequency of the probing field, as has been investigated in detail in the off-resonant configuration~\cite{Hofferberth2007}. In this section, the position and strengths of the expected resonances are discussed.

\subsection{Effect of the dressing field}
In the following, we denote as $(x,y,z)$ the axes of the laboratory frame, $z$ being the vertical axis. In the experiment, the adiabatic potential is created by a strong RF field of frequency $\omega_1$ and Rabi frequency $\Omega_1$, linearly polarised along $y$, which dresses the atoms placed in an inhomogeneous magnetic field $B(\mathbf{r})$. The direction of this static magnetic field at position $\mathbf{r}$ is denoted as $Z$. The two important parameters which determine the resulting adiabatic potentials are the RF detuning $\delta(\mathbf{r}) = \omega_1 - \omega_L(\mathbf{r})$ from the local Larmor frequency and the local effective RF coupling $\Omega_1^{\mbox{\scriptsize eff}}(\mathbf{r})$. The latter depends on the angle $\theta(\mathbf{r})$ between the RF polarisation $y$ and the static magnetic field direction $Z$, being zero for collinear fields: $\Omega_1^{\mbox{\scriptsize eff}}(\mathbf{r})=\Omega_1\sin\theta(\mathbf{r})$. Hence, the static magnetic field geometry is the skeleton of the adiabatic potentials.

The static magnetic field $B(\mathbf{r})$ is that of a compact Ioffe-Pritchard trap~\cite{Esslinger1998} characterised by the bias field $B_0 = 1.8$\,G at the centre, the radial gradient $b'=223$\,G$\cdot$cm$^{-1}$ and the longitudinal curvature $b''=250$\,G$\cdot$cm$^{-2}$ along the trap axis $x$. We define as $\omega_0= |g_F|\mu_B B_0/\hbar$ the Larmor frequency at the trap centre and as $\alpha = |g_F|\mu_B b'/\hbar$ the magnetic gradient in units of frequency, where $g_F$ is the Land\'e factor, $\sigma(g_F)=g_F/|g_F|$ its sign and $\mu_B$ is the Bohr magneton. The typical length-scales in this magnetic trap are $L=\sqrt{B_0/b''}=850~\mu$m and $R=B_0/b'=\omega_0/\alpha=81~\mu$m. The Larmor frequency at position $\mathbf{r}$ is then approximately given by $\omega_L(\mathbf{r}) = \omega_0\sqrt{1+x^2/L^2+(y^2+z^2)/R^2}$ in the vicinity of the magnetic trap centre. On the other hand, the local effective coupling to the RF field is approximately $\Omega_1^{\mbox{\scriptsize eff}}(\mathbf{r}) = \Omega_1\sqrt{1-\frac{\alpha^2y^2}{\omega_L(\mathbf{r})^2}}$.

As a result of the near resonant dressing field, the energy levels of the atom plus RF field system are grouped into manifolds separated by an energy $\hbar \omega_1$. Inside a manifold, the energy spacing of the $2F+1$ levels is denoted $\hbar\Omega(\mathbf{r})$. The adiabatic potential $V_1^{m_F}(\mathbf{r}) = \sigma(g_F)\, m_F \hbar\Omega(\mathbf{r})$ then varies in space through the spatial variation of both the detuning $\delta(\mathbf{r})$ and the effective RF coupling $\Omega_1^{\mbox{\scriptsize eff}}(\mathbf{r})$~\cite{Lesanovsky2006}. Given the static magnetic field geometry, the adiabatic potential for a given $F,m_F$ adiabatic state reads:
\begin{eqnarray}
V_1^{m_F}(\mathbf{r}) &=& \sigma(g_F)\, m_F \hbar\Omega(\mathbf{r})\mbox{, where} \label{eq:potential}\\
\Omega(\mathbf{r})&=& \sqrt{\delta(\mathbf{r})^2 + \Omega_1^2\left(1-\frac{\alpha^2y^2}{\omega_L(\mathbf{r})^2}\right)}. \nonumber
\end{eqnarray}
This expression corresponds to the eigenenergies at point $\mathbf{r}$ of the Hamiltonian $H_1 = -\delta(\mathbf{r}) F_Z + \Omega_1^{\mbox{\scriptsize eff}}(\mathbf{r}) F_X$, obtained after applying the rotating wave approximation (RWA) to the RF-atom coupling, which is valid when $\Omega_1 \ll \omega_1$ and $|\delta(\mathbf{r})| \ll \omega_1$. Here, $X$ is direction orthogonal to the static field direction $Z$, in the $y-Z$ plane. The total potential energy is $V^{m_F}(\mathbf{r}) = V_1^{m_F}(\mathbf{r}) + Mgz$, where $g$ is the gravitational acceleration and $M$ the atomic mass.

\begin{figure}[t]
 \begin{center}
   \includegraphics[width=0.9\columnwidth]{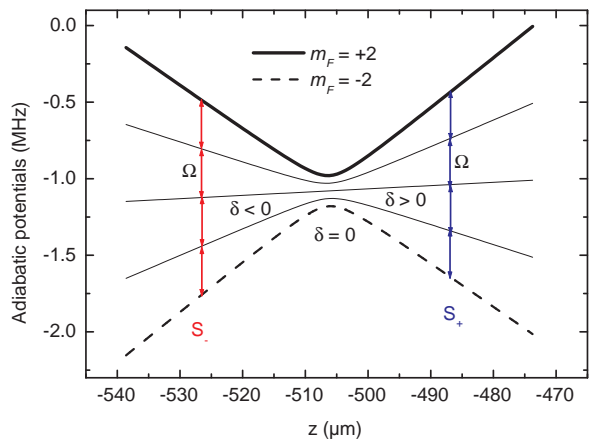}
 \end{center}
\caption{
Adiabatic potential energy $(V_1^{m_F}(z) + Mgz)/h$ for the $F=2$ ground state of $^{87}$Rb in our magnetic field configuration, in units of frequency of the dressed states $m_F=2,...-2$, plotted along the vertical direction. The dressing frequency is set to $\omega_1 = 2\pi\times 8$~MHz with a Rabi frequency of $\Omega_1 = 2\pi\times 50$~kHz, resulting in a dressed trap centred at position $z = -506~\mu$m. The two extreme dressed levels are emphasised. The origin $z=0$ corresponds to the static magnetic field minimum, and gravity is taken into account. The probing field $\omega_2$ induces transitions between dressed states at a position $z$ such that $\Omega(z) = |\omega_2 - n\omega_1|, n \in \mathbb{N}$. These positions correspond to two points $z_{\pm}$ belonging to two surfaces $S_{\pm}$, defined respectively by $\delta(\mathbf{r})=\pm\sqrt{(n\omega_1-\omega_2)^2 - \Omega_1^{\mbox{\scriptsize eff}}(\mathbf{r})^2}$, on both sides of the trapping surface $S_0$ defined by $\delta(\mathbf{r})=0$.}
\label{fig:levels}
\end{figure}

The potential minimum for the extreme adiabatic state $m_F = \sigma(g_F)\,F$ is located on the isomagnetic surface $S_0$ defined by $\delta(\mathbf{r}) = 0$ or equivalently $\omega_L(\mathbf{r}) = \omega_1$~\cite{note1}. Schematically, this surface is an elongated ellipsoid, cylindrically symmetric around $x$ with a radius $r_0=\sqrt{\omega_1^2 - \omega_0^2}/\alpha$ and a long axis $x_0=r_0 L/R$. On the surface $S_0$, $\omega_L(\mathbf{r}) = \omega_1$ and the RF coupling only depends on the coordinate $y$: $\Omega_1^{\mbox{\scriptsize eff}}(y) = \Omega_1\sqrt{1-\frac{\alpha^2y^2}{\omega_1^2}}$. The total potential energy on $S_0$ then reads $F\hbar \Omega_1^{\mbox{\scriptsize eff}}(y) + Mgz$. It varies through the RF shift between $F\hbar\Omega_1\sqrt{1-\frac{\alpha^2r_0^2}{\omega_1^2}}=F\hbar\frac{\omega_0}{\omega_1}\Omega_1$ at $y=\pm r_0$ on the $y$ axis and $F\hbar\Omega_1$ in the $y=0$ plane. On the other hand, the gravitational shift also varies as a function of $z$ between $-Mgr_0$ at the bottom $(z=-r_0)$ and 0 at the equator $(z=0)$. Depending on the relative strength of the gravitational shift $Mgr_0$ with respect to the RF shift $F \hbar\Omega_1(1-\frac{\omega_0}{\omega_1})$, there are either two minima at the equator at $z=0$ and $y=\pm r_0$ \cite{Schumm2005} or a single minimum at the bottom of this surface $z=-r_0$ \cite{Morizot2007}. With the parameters of our experiment, we are always in the second situation \cite{Colombe2004}. At the minimum, where the atomic density is largest, the RF coupling reaches its maximum value $\Omega_1$. Away from the minimum, it decreases down to a value depending on the energy and on the RF frequency through the ellipsoid radius $r_0$. At a given energy above the potential minimum, the coupling strength reduction is more pronounced for smaller RF frequencies $\omega_1$. For example, at the operational RF frequency of 8\,MHz the reduction is less than 20\% at energies below $10\,\mu$K. By contrast, at 3\,MHz, the reduction remains below 7\% at an energy corresponding to $1~\mu$K, and reaches 50\% for 10\,$\mu$K. Nevertheless, the effective coupling $\Omega_1^{\mbox{\scriptsize eff}}$ never vanishes over $S_0$. If necessary, the coupling inhomogeneity could be reduced by increasing the bias field $B_0$.

In the following, we concentrate on the case of rubidium 87 in the state $F=2$. A plot of the adiabatic potentials around $z=-r_0$ as a function of the vertical coordinate $z$ in the laboratory frame is given in figure~\ref{fig:levels} for a typical RF frequency $\omega_1 = 2\pi\times 8$~MHz and for the parameters of our Ioffe-Pritchard magnetic trap. The resulting trap in the dressed state $m_F=F=2$ is very anisotropic, with oscillation frequencies near the trap centre of the order of 1~kHz along $z$ and of 5 to 20~Hz in the horizontal directions. In this way, the atom cloud has the shape of a \textit{cr\^epe}~\cite{Colombe2004}.

\subsection{Coupling to the probing field: resonance frequencies}

A second RF field of frequency $\omega_2$ and Rabi frequency $\Omega_2$ is applied to the dressed atoms. The polarisation of this second field is also linear along $y$, and again its orientation relative to the static magnetic field slightly varies across the dressed trap. As a result, both polarisations $X$ and $Z$ are present and can couple to the atomic spin. In the following, we neglect the position dependence of $\Omega_2$ and concentrate in each case on the polarisation component having a significant coupling to the atomic spin. $\Omega_2$ is typically a factor 100 times smaller than the dressing Rabi frequency $\Omega_1$. This second field allows a coupling between dressed states $m_F$ and $m'_F$ with $|m_F-m'_F|=1$: see the arrows in figure~\ref{fig:levels}. This coupling occurs either within the same manifold or between different manifolds. The expected resonant frequencies for this coupling at position $\mathbf{r}$ are thus $\omega_2 = n\omega_1 \pm \Omega(\mathbf{r})$, $n$ being an integer~\cite{Hofferberth2007}.

\begin{figure}[t]
 \begin{center}
   \includegraphics[width=0.8\columnwidth]{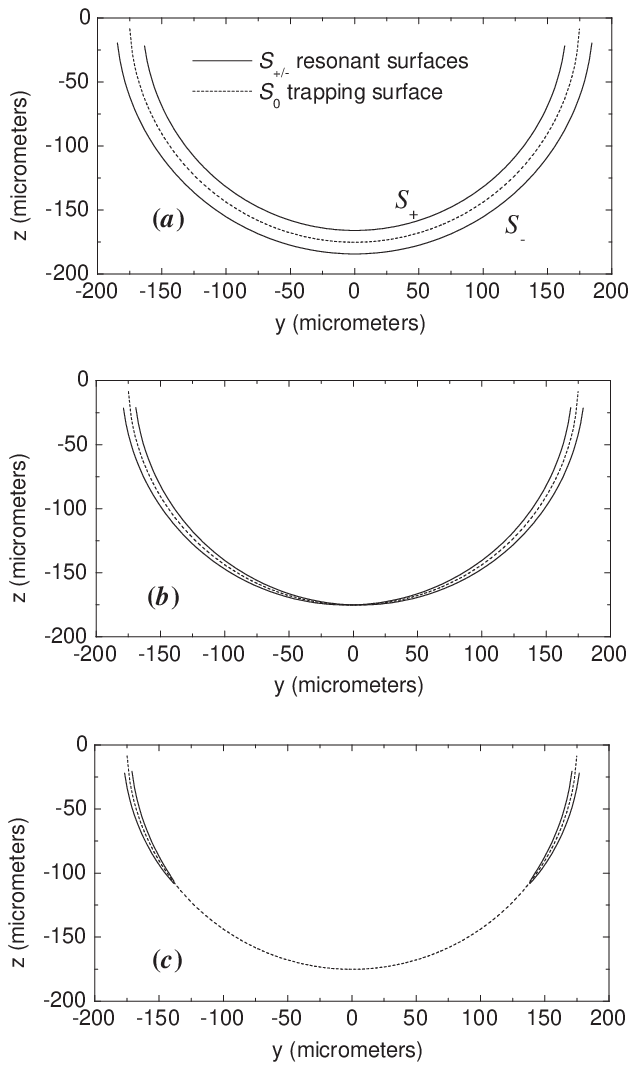}
   
   \includegraphics[width=0.7\columnwidth]{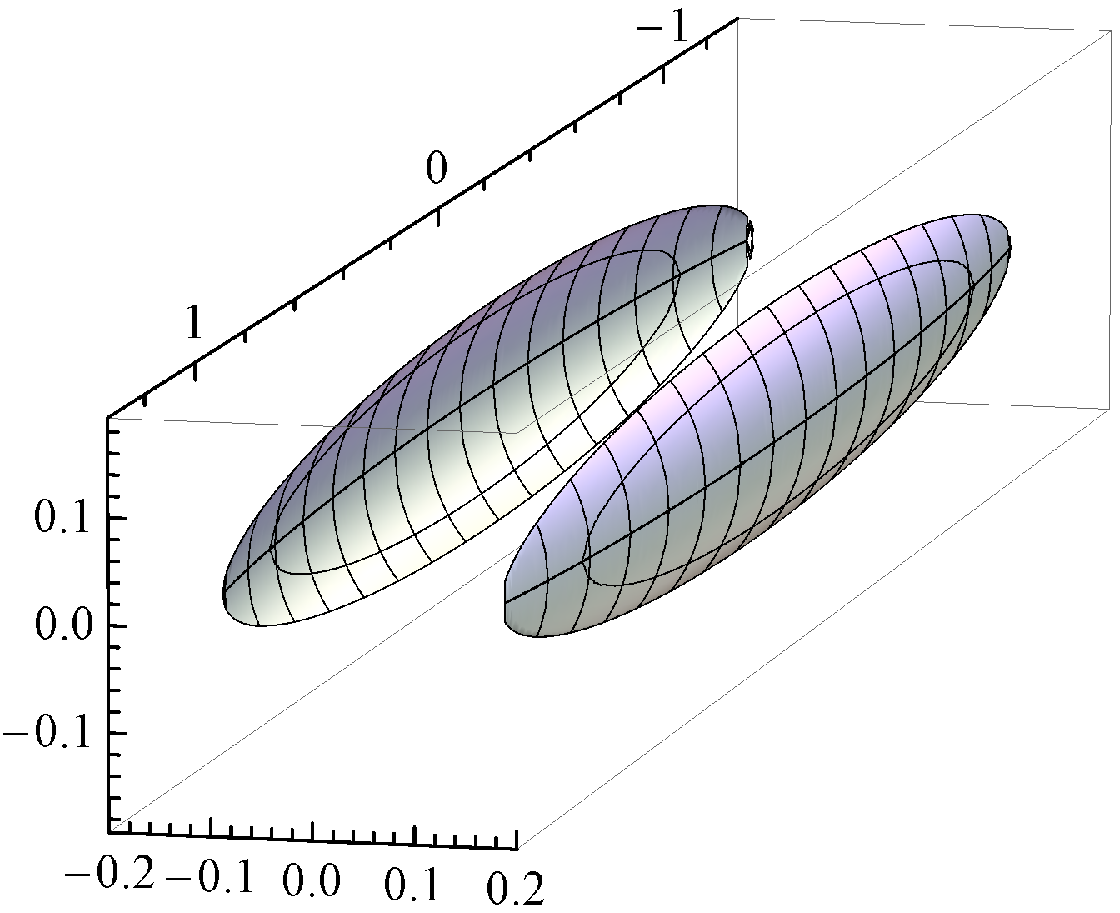}\hspace{-5mm}{\footnotesize $(d)$}
 \end{center}
\caption{Cut of the resonant surfaces $S_{\pm}$ in the plane $x=0$ in the three cases discussed in the paper. $(a)$: $|\omega_2 - n\omega_1|=2\Omega_1>\Omega_1$, $(b)$: $|\omega_2 - n\omega_1|=\Omega_1$, $(c)$: $|\omega_2 - n\omega_1|=0.7\Omega_1$, \textit{i.e.} $\Omega_1\omega_0/\omega_1<|\omega_2 - n\omega_1|<\Omega_1$. The isomagnetic surface $S_0$ to which the atoms are trapped is also represented. $\omega_1$ is set to $2\pi\times 3$\,MHz to show the potentials clearly and $\Omega_1$ is set to $2\pi\times 75$\,kHz. $(d)$: Corresponding outer resonant surface $S_-$ in a 3D representation in the case $(c)$.}
\label{fig:res_surfaces}
\end{figure}

In the `resonant configuration', the points $\mathbf{r}$ in space where the probing field is resonant, \textit{i.e.} where the condition $\Omega(\mathbf{r}) = |\omega_2 - n\omega_1|$ is satisfied, consist of two surfaces $S_+$ and $S_-$ on both sides of the trapping surface $S_0$ (see figure~\ref{fig:res_surfaces}$a$). The detuning of the local Larmor frequency with respect to $\omega_1$ has an opposite sign for these two probing surfaces, $\delta(\mathbf{r})=\pm\sqrt{(n\omega_1-\omega_2)^2 - \Omega_1^{\mbox{\scriptsize eff}}(\mathbf{r})^2}$ for the surface $S_{\pm}$ respectively.

The shape of the resonant surfaces $S_{\pm}$ strongly depends on the value of the probing frequency $\omega_2$ around $n\omega_1$. First, let's remark that the resonance can be reached at point $\mathbf{r}$ for probing frequencies $\omega_2$ close to $n\omega_1$ only if $|\omega_2 - n\omega_1|>\Omega_1^{\mbox{\scriptsize eff}}(\mathbf{r})$. Three cases can then be distinguished. First, for frequencies $\omega_2$ satisfying $|\omega_2 - n\omega_1|>\Omega_1$, the resonant surfaces are very close to isomagnetic ellipsoids, see figure~\ref{fig:res_surfaces}$a$. The atoms need a minimum non zero energy to reach these surfaces, and the point where the necessary energy is the lowest lies at the bottom of the external surface $S_-$. Second, for $|\omega_2 - n\omega_1|=\Omega_1$, the two resonant surfaces merge at the trapping surface $S_0$ all over the vertical plane $y=0$, in particular at the trap bottom, and all atoms may reach this point and be lost, see figure~\ref{fig:res_surfaces}$b$. Third, for $|\omega_2 - n\omega_1|<\Omega_1$, resonant surfaces still exist due to the spatial dependence of $\Omega_1^{\mbox{\scriptsize eff}}$, but each of them is split into two curved surfaces centred around the $y$ axis, see figure~\ref{fig:res_surfaces}$c$ and \ref{fig:res_surfaces}$d$. Again a minimum energy is necessary to reach the lowest resonant point. Finally, for $|\omega_2 - n\omega_1|$ equal to $\Omega_1\omega_0/\omega_1$, the resonant surfaces collapse to two points at the equator and for lower values of $|\omega_2 - n\omega_1|$ no atoms are lost.

\begin{figure}[t]
 \begin{center}
   \includegraphics[width=0.9\columnwidth]{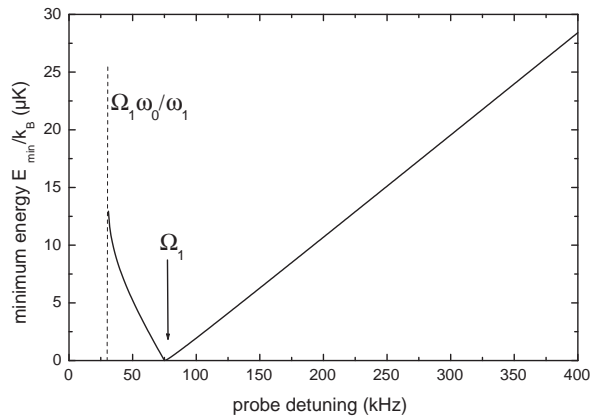}
 \end{center}
\caption{Minimum energy required for an atom to leave the dressed trap by reaching a resonant surface, as a function of the probe detuning $|\omega_2 - n\omega_1|$. The parameters are: $\omega_1=2\pi\times 3$\,MHz and $\Omega_1=2\pi\times 75$\,kHz.}
\label{fig:minrespot}
\end{figure}

In summary, the energy in the dressed trap necessary for an atom to escape through a point resonant with the probing field depends on the probe detuning to $n\omega_1$. Figure~\ref{fig:minrespot} gives, as a function of $|\omega_2 - n\omega_1|$, the minimum energy $E_{\mbox{\scriptsize min}}(\omega_2)$ with respect to the trap bottom for an atom to escape, for a dressing frequency of $\omega_1/2\pi=3$\,MHz and a Rabi frequency of $\Omega_1/2\pi=75$\,kHz. The three regimes discussed above are clearly visible. The final shape of the recorded spectrum can be inferred from this minimum energy. If, over the duration of the experiment, the RF weak probe couples atoms efficiently out of the dressed trap when a resonant surface is reached, all the atoms with an external energy $E$ larger than $E_{\mbox{\scriptsize min}}(\omega_2)$ should be lost. Hence, the expected fraction of remaining atoms in the trap $f_{\mbox{\scriptsize rem}}(\omega_2)$ is a function of $E_{\mbox{\scriptsize min}}(\omega_2)$, namely
\begin{equation}
f_{\mbox{\scriptsize rem}} = \int_0^{E_{\mbox{\scriptsize min}}(\omega_2)} p(E) \, dE
\label{eq:frem}
\end{equation}
where $p(E)$ is the energy distribution in the dressed trap, and the zero energy level for $E$ is taken at the trap bottom. In particular, we expect $f_{\mbox{\scriptsize rem}}(n\omega_1\pm \Omega_1)=0$ and an asymmetric shape of the spectrum around this value.

The exact shape of $p(E)$ is related to the exact energy potential. Its expression is not analytical, as the trap is far from being harmonic. However, we propose as a simple ansatz an exponential expression for $p$ of the form $p(E)=(k_BT)^{-1}\exp(-E/k_BT)$ and compare the simple prediction $f_{\mbox{\scriptsize rem}}(\omega_2) = 1 - \exp[-E_{\mbox{\scriptsize min}}(\omega_2)/k_BT]$ to the experimental data, see section~\ref{Sec:Results}.

\subsection{Coupling to the probing field: coupling strengths}

In the previous section, the two resonant surfaces $S_{\pm}$ at which the transitions $\omega_2 = n\omega_1 \pm \Omega(\mathbf{r})$ can occur have been presented. Let us point out that the strength of these different transitions is not the same, and depends on position in the dressed trap. This is true even if we assume a homogeneous Rabi frequency $\Omega_2$, which we do in the following for simplicity. For example for $n=1$, the coupling strength at position $\mathbf{r}$ is $\frac{\Omega_2}{2}\left(1\mp \frac{\delta(\mathbf{r})}{\Omega(\mathbf{r})}\right)$ for the transitions at frequency $\omega_2 = \omega_1 \pm  \Omega(\mathbf{r})$, respectively~\cite{Garrido2006}. The coupling strength can be interpreted in terms of the number of photons $n$ from the dressing field necessary for the transition to occur. If we consider detunings $\delta(\mathbf{r})$ which are large compared to the dressing amplitude $\Omega_1$ (such that $\Omega(\mathbf{r})\simeq|\delta(\mathbf{r})|=|\omega_1 - \omega_L(\mathbf{r})|$), then the atoms are far from the trapping surface $S_0$. In this case, the two RF couplings become independent and the probing field essentially couples undressed states together. As a result, the coupling is of order $\Omega_2$ for the single photon process at frequency $\omega_2 \simeq \omega_L(\mathbf{r})$, and drops to a vanishing value $\displaystyle\Omega_2 \frac{\Omega_1^{\mbox{\scriptsize eff}}(\mathbf{r})^2}{4\delta^2}$ for the three photon process at frequency $\omega_2 \simeq 2 \omega_1 - \omega_L(\mathbf{r})$. Hence, transitions to an untrapped state occur essentially through one of the two surfaces resonant with the probing field: the inner surface $S_+$, closer to the magnetic field minimum, for the $\omega_2 = \omega_1 - \Omega(\mathbf{r})$ transition, and the outer surface $S_-$, below the trapped atoms, for the $\omega_2 = \omega_1 +  \Omega(\mathbf{r})$ transition.

Let us now focus on the low frequency transition $n=0$, at $\omega_2 = \Omega(\mathbf{r})$. In contrast to the previous case, this transition presents (as we shall see below) a symmetric coupling $\Omega_1^{\mbox{\scriptsize eff}}(\mathbf{r})\Omega_2/\omega_2$ at the two resonant surfaces $S_{\pm}$, between which the trapped atoms lie. This has some importance in the implementation of evaporative cooling in the dressed trap.

As this transition was not discussed in Ref.\cite{Garrido2006}, we will briefly present its origin. First, we remark that the RF probe field should be $\pi$-polarised, meaning that the RF field component \textit{parallel} to the static magnetic field direction $Z$ is responsible for the transition. This can be understood by considering the transition induced by the probe away from the trapping surface $S_0$, where the two RF fields decouple and the resonant condition is simply $\omega_1 \pm \omega_2 = \omega_L(\mathbf{r})$: the probe and the \textit{circularly} polarized dressing field $\omega_1$ induce a two-photon transition, that will change $m_F$ by 1 unit if the probe field is $\pi$-polarized. The Hamiltonian describing the spin evolution at position $\mathbf{r}$ in the presence of the two RF fields then reads
\begin{equation}
  H(t) = \omega_L(\mathbf{r}) F_Z + 2 \Omega_1^{\mbox{\scriptsize eff}}(\mathbf{r}) F_X \cos\omega_1 t + 2 \Omega_2 F_Z \cos\omega_2 t 
\end{equation}
where $Z$ is the direction of quantization imposed by the static magnetic field and $X$ is the effective polarisation of the dressing RF field $\omega_1$ of Rabi frequency $\Omega_1^{\mbox{\scriptsize eff}}(\mathbf{r})$. To deal with this double frequency RF coupling, we apply a rotation transformation to the Hamiltonian in the spirit of Ref.\cite{Garrido2006}, at an angle $\theta(t) = (\omega_1 t + \frac{2\Omega_2}{\omega_2}\sin\omega_2t)$ around the $Z$ axis. The idea of this transformation is that the term in $\dot{\theta} F_Z$ is chosen to cancel the new term $2 \Omega_2 F_Z \cos\omega_2 t$. After the rotation, product terms of the form $\cos\omega_1 t \cos\left(\omega_1t+ \frac{2\Omega_2}{\omega_2}\sin\omega_2t\right)$ appear. Since $\Omega_2 \ll \omega_2$, these terms can be developed as products of sine and cosine functions. Finally, application of the rotating wave approximation allows the elimination of fast rotating terms (rotating at about $2\omega_1$). The Hamiltonian in the rotated frame finally simplifies to
\begin{equation}
  H'(t) \simeq -\delta(\mathbf{r}) F_Z + \Omega_1^{\mbox{\scriptsize eff}}(\mathbf{r}) F_X - \frac{2\Omega_1^{\mbox{\scriptsize eff}}(\mathbf{r})\Omega_2}{\omega_2} \sin\omega_2t F_Y.
\end{equation}
This consists of the Hamiltonian $H_1=-\delta(\mathbf{r}) F_Z + \Omega_1^{\mbox{\scriptsize eff}}(\mathbf{r}) F_X$ for the state dressed by the main RF field, to which a transverse RF coupling of frequency $\omega_2$ and effective amplitude $ \frac{\Omega_1^{\mbox{\scriptsize eff}}(\mathbf{r})\Omega_2}{\omega_2}$ is added. This coupling allows transitions at a frequency $\omega_2 = \Omega(\mathbf{r})$. Again, the condition is fulfilled at the two resonant surfaces $S_{\pm}$, but now with the same coupling amplitude at two points symmetric with respect to the surface $S_0$, in contrast to the $n=1$ case.

\section{Experimental results}
In this section, experimental results of RF spectroscopy of a RF-dressed trap are presented. The results are obtained in a dressed Ioffe-Pritchard trap in the resonant configuration.
\subsection{Experimental procedure}
\label{Sec:Experiment}
The loading procedure into the adiabatic potential was described in detail in a previous paper~\cite{Morizot2008}. Rubidium 87 atoms are initially confined in their internal state $F=2, m_F = 2$ in a static magnetic trap and evaporatively cooled to a temperature $T_0$. The dressed trap is loaded into the dressed state $m_F = 2$ with a 500~ms RF ramp from 800~kHz to the final value of $\omega_1$, ranging between 3 and 8~MHz. At the end of this loading stage, a second RF field with a fixed frequency $\omega_2$ and a Rabi frequency $\Omega_2$ is switched on for 2 seconds, to perform the spectroscopy of the adiabatic potential. The Rabi frequency $\Omega_1$ of the dressing field is kept constant during the loading stage and the probing stage~\cite{note2}. The RF fields are produced by two identical antennas (9 windings, 10~mm in diameter) placed face to face across the vacuum chamber, at a distance of 17~mm from the atomic cloud. At the input of each antenna, the RF amplitude lies between 160 and 640\,mA$_{pp}$ for the dressing field, and between 0.8 and 5\,mA$_{pp}$ for the probing field. Finally, the atom number and the optical density are recorded after a time of flight of 7 or 10~ms, and a spectrum is recorded as a function of the probing frequency $\omega_2$. This procedure is repeated for different values of $\Omega_1$, $\Omega_2$ and $\omega_1$. As discussed in the previous section, resonances are expected at frequencies around $\Omega_1$, $\omega_1 - \Omega_1$, $\omega_1 + \Omega_1$, $2\omega_1-\Omega_1$, etc. \cite{Hofferberth2007}.

\begin{figure}[t]
 \begin{center}
   \includegraphics[width=0.9\columnwidth]{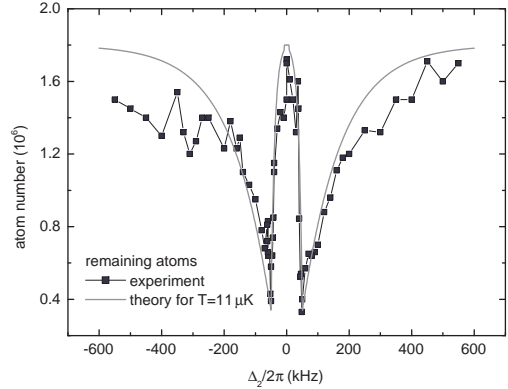}
 \end{center}
\caption{Spectroscopy of the dressed trap, around the dressing frequency $\omega_1 = 2\pi\times 8$\,MHz. The number of atoms remaining after the
application of the probe field is plotted (squares) as a function of the detuning $\Delta_2 = \omega_2-\omega_1$ to the frequency $\omega_1$. The dressing amplitude is 400\,mA$_{pp}$ at the input of the RF antenna, corresponding to a Rabi frequency $\Omega_1=2\pi\times 50$\,kHz as deduced from the spectrum. The probing amplitude is 1.6\,mA$_{pp}$, smaller by a factor 250. The calculated number of remaining atoms for $T=11~\mu$K is plotted as a full grey line, see text.}
\label{fig:2peaks}
\end{figure}

\subsection{Experimental spectra}
\label{Sec:Results}

A typical spectrum recorded around $\omega_1 = 2\pi\times 8$\,MHz is plotted in figure~\ref{fig:2peaks}. For this data set, the dressing amplitude is set to 400\,mA$_{pp}$ while the probing amplitude is 1.6\,mA$_{pp}$. The two resonances at $\omega_1\pm\Omega_1$ are visible, broadened by the energy distribution in the trap at temperature $T$. For probing frequencies between $\omega_1-\Omega_1$ and $\omega_1+\Omega_1$, the atomic loss rate rapidly decreases to zero, and no transition is observed at a frequency $\omega_2=\omega_1$, as expected from the previous discussion. At $\omega_2=\omega_1\pm\Omega_1$, the probing frequency is resonant with the bottom of the adiabatic potential and almost all the atoms are lost (80\%\ for this particular choice of the probe coupling $\Omega_2$).

Away from this value, the atoms satisfying $\Omega(\mathbf{r}) > |\omega_2-\omega_1|$ are lost. As expected, the observed line has an asymmetric shape, with a dip at $\omega_2=\omega_1\pm\Omega_1$ and approximately exponential outer wings with a width related to the temperature. Indeed, in a harmonic trap, the lineshape should be exponential, with a $1/e$-width $k_B T/(2\hbar)$ and a threshold at an energy corresponding to the trap bottom. Using an exponential trial function $p(E)=(k_BT)^{-1}\exp(-E/k_BT)$ for the energy distribution, the resulting expression for $f_{\mbox{\scriptsize rem}}$ derived from Eq.(\ref{eq:frem}), and the temperature of $T=11~\mu$K deduced from an independent time-of-flight measurement, we obtain the predicted spectrum plotted with a grey line in figure~\ref{fig:2peaks}. $f_{\mbox{\scriptsize rem}}$ is normalised to total the atom number, and the limited maximum contrast of 80\% of atom loss for 2~s at resonance is also taken into account. The experimental spectrum is in good agreement with the predicted lineshape, indicating that the exponential model used to estimate the energy distribution is indeed sufficient.

\begin{figure}[t]
 \begin{center}
   \includegraphics[width=0.9\columnwidth]{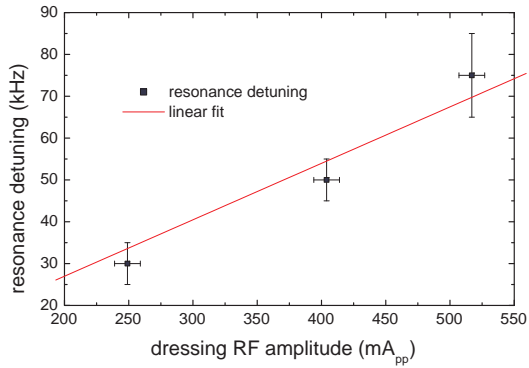}
 \end{center}
\caption{Rabi frequency $\Omega_1$, as deduced from the position of the edge of the line, plotted as a function of the dressing RF amplitude. RF dressing frequency: $\omega_1=2\pi\times 8$\,MHz.}
\label{fig:Rabi_vs_amplitude}
\end{figure}

The sharp edge of the spectrum allows a good determination of the dressing Rabi frequency $\Omega_1$ at the bottom of the adiabatic potential, equal to 50\,kHz in the spectrum of figure~\ref{fig:2peaks}. The determination is more accurate when a smaller probe Rabi frequency is used. A larger amplitude $\Omega_2$ leads to broadening of the resonance line. Experimentally, an amplitude of 1.6 to 3.2\,mA$_{pp}$ was found to be a good compromise between a good probing efficiency (80\% of atom loss after 2~s on resonance) and a negligible line broadening. The RF dressing amplitude was varied with a RF attenuator and the position of the peak atom loss was recorded to check the validity of the determination of $\Omega_1$ by this technique. The result is plotted on figure~\ref{fig:Rabi_vs_amplitude}. The detuning at the maximum atom loss is expected to be directly proportional to the RF amplitude. At a dressing frequency $\omega_1=2\pi\times 8$\,MHz, the deduced Rabi frequency is $\Omega_1 = \eta I_1$, with a scaling factor $\eta = 2\pi\times 135$\,kHz/A$_{pp}$, $I_1$ being the RF amplitude in amperes peak-to-peak. This is in agreement with the value of 130\,kHz/A$_{pp}$ calculated from the geometric characteristics of the antenna.

\begin{figure}[t]
 \begin{center}
   \includegraphics[width=0.9\columnwidth]{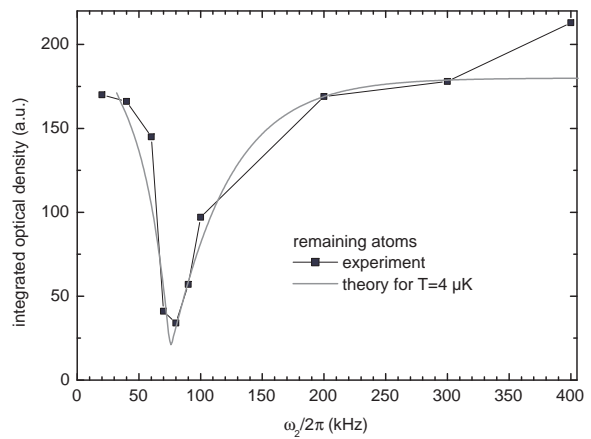}
 \end{center}
\caption{Spectroscopy of the dressed trap, around the dressing amplitude $\Omega_1 = 2\pi\times 75$\,kHz, for a dressing frequency $\omega_1=2\pi\times 3$~MHz. The horizontal axis indicates the probe frequency $\omega_2$. The dressing amplitude is 580\,mA$_{pp}$ at the input of the RF antenna, corresponding to a Rabi frequency $\Omega_1=2\pi\times 75$\,kHz (see text). The probing amplitude is 9\,mA$_{pp}$ at the input of the second antenna, but only the component parallel to $Z$ couples to the atoms (see text). The calculated number of remaining atoms for $T=4~\mu$K is plotted as a full grey line.}
\label{fig:1peak}
\end{figure}

The determination of $\Omega_1$ from the spectra around $\omega_2=\omega_1\pm\Omega_1$ is also confirmed by the data recorded at low probing frequency. Figure~\ref{fig:1peak} presents a typical spectrum at frequency $\omega_2 \simeq \Omega_1$. We recover a Rabi frequency of 75\,kHz at 3\,MHz of dressing frequency. Again, the asymmetric shape is due to energy distribution of the atoms confined in the dressed trap. The overall shape of the spectrum is in reasonable agreement with the calculated spectrum for a temperature of $4~\mu$K.

\subsection{Evaporative cooling}
\begin{figure}[t]
 \begin{center}
   \includegraphics[width=0.9\columnwidth]{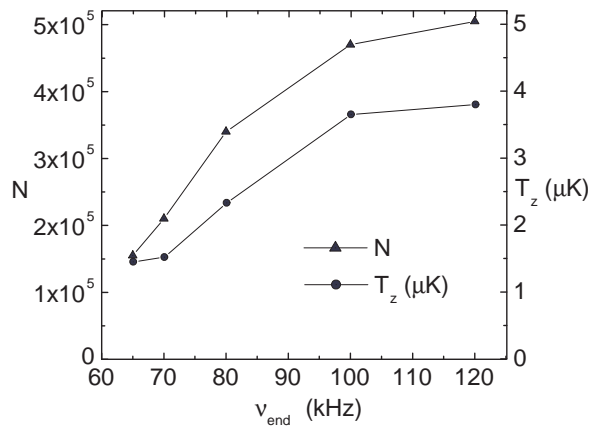}
 \end{center}
\caption{Final atom number $N$ and kinetic temperature $T_z$, deduced from the vertical size of the cloud after a time-of-flight expansion, after an evaporation ramp of the second RF field between 600\,kHz and $\nu_{\mbox{\scriptsize end}}$.}
\label{fig:evaporation}
\end{figure}

Garrido Alzar \textit{et al.} \cite{Garrido2006} proposed to use this second RF in this trap to force evaporation and cool the atoms. Indeed, the energy sensitivity of the atom loss induced by the second RF can be used to selectively remove the most energetic atoms from the trap. We implemented this forced evaporative cooling inside the RF-dressed trap. For this purpose, the second RF was scanned from an initial frequency away from the resonance down to a frequency close to the resonance at the trap bottom. The experiments where performed with a dressing RF amplitude of $\Omega_1 = 2\pi\times 50$\,kHz, at $\omega_1 = 2\pi\times 8$\,MHz. To optimize the symmetry of the coupling through the two resonant surfaces $S_{\pm}$, the low frequency resonance was preferred. The frequency of the weak field $\omega_2/2\pi$ was lowered from 600~kHz to a final value $\nu_{\mbox{\scriptsize end}}$. We observed a decrease of the cloud temperature and the atom number as $\nu_{\mbox{\scriptsize end}}$ is lowered, see figure~\ref{fig:evaporation}. The selective out-coupling of the atoms with a large vertical energy $E_z$ works as expected around the low frequency resonance, whereas almost no effect was obtained when scanning the frequency around $\omega_1$, where the out-coupling is asymmetric. In our dressed trap, the phase space density increases only slightly, due to a poor thermalisation efficiency with the low horizontal frequencies (5~Hz $\times$ 20~Hz). With larger horizontal frequencies in the dressed trap, as is the case in a dressed quadrupole trap for example~\cite{Morizot2007}, and thus with a higher collision rate, thermalisation in the trap would lead to an increase of the phase space density~\cite{Luiten1996}, as shown by molecular dynamics simulations. From the point of view of the out-coupling efficiency, our results show that evaporative cooling is feasible in a resonant dressed trap in the same conditions as in a conventional magnetic trap.

\section{Conclusion and prospects}
In this paper, we have demonstrated RF spectroscopy in a RF dressed trap operated in a regime where the trapped atoms are resonant with the dressing RF source. Thus we use two RF fields: a strong dressing RF source, and a second weak probing field which couples the atoms out of the dressed trap. Spectra are recorded either around the dressing RF frequency (a few MHz), or at low frequencies near the Rabi frequency (a few tens of kHz). The experimental data fit well with a simple model based on the minimum energy for a trapped atom to be outcoupled by the weak probe. The low frequency transition is used to selectively evaporate the hotter atoms from the dressed trap. A temperature decrease is demonstrated in the transverse direction of the dressed trap together with an atomic population decrease. The resulting phase space density increase is modest at the present stage of the experiment since the very low motional frequencies of this anisotropic trap imply a low collision rate and thus a poor thermalisation.

The results of this paper emphasise the interest in utilizing resonant RF fields to realise unusual atomic potentials and show that we can apply evaporative cooling inside these potentials. Unlike light based traps, a resonant RF-based trap is not affected by spontaneous emission. On the other hand it is sensitive to the noise of the RF source. However, heating, and the ensuing losses, can be kept at very low values provided there is a good control of the parameters of the source~\cite{Morizot2008}. Finally, we conclude that with reasonable trap frequencies (in all directions) it should be possible to have an efficient RF evaporative cooling of an atomic population in a RF trap.

\acknowledgments
We thank V.~Dini for assistance in data processing. This work was supported by the R\'egion Ile-de-France (contract number E1213 and IFRAF), by the PPF `Manipulation d'atomes froids par des lasers de puissance', by the European Community through the Marie Curie Training Network `Atom Chips' under contract number MRTN-CT-2003-505032, and by Leverhulme Trust. B.\,M.\,G. thanks CNRS for support. Laboratoire de physique des lasers is UMR 7538
of CNRS and Paris 13 University. LPL is member of the Institut Francilien de Recherche sur les Atomes Froids (IFRAF).


\begin{thebibliography}{40}
\bibitem{note1} Strictly speaking, the minimum is shifted away from the isomagnetic surface $S_0$ by gravity along $z$ and by the RF coupling variation along $y$. However, these shifts are negligible if $\Omega_1/\omega_1 \ll m_F\hbar\alpha/(Mg)$ and $\Omega_1^2/\omega_1^2 \ll 1$, respectively. With the figures of our experiments, the relative correction on the position is less than 0.1\% of $r_0$.
\bibitem{note2} For the lowest value of the Rabi frequency $\Omega_1 = 2\pi\times 30$~kHz tested here, the RF power was larger during the loading ramp to limit losses, and was later reduced to the desired value for the probing stage.
\end{thebibliography}
\end{document}